\documentclass[conference]{IEEEtran}
\usepackage{color,graphicx}
\usepackage[font=footnotesize, labelfont={sf,bf}, margin=0cm]{caption}
\usepackage{url}
\usepackage{comment}
\usepackage{amsmath}
\usepackage{algpseudocode}
\usepackage{algorithmicx}
\usepackage{algorithm}
\usepackage{epstopdf}
\usepackage{epsfig}
\usepackage[T1]{fontenc}
\usepackage{txfonts}
\usepackage{cite}
\usepackage{array,multirow,graphicx}

\newcommand{\la}{\leftarrow}
\newcommand{\ra}{\rightarrow}
\newcommand{\bs}{\backslash}

\newcommand{\argmax}[1]{\underset{#1}{\operatorname{arg}\,\operatorname{max}}\;}

\newcommand{\Rmnum}[1]{\expandafter\@slowromancap\romannumeral #1@}
\begin{document}
%\title{Exploiting Full Duplex Small Cell Networks with D2D for Improving Network Capacity}
\title{On Improving Capacity of Full-Duplex Small Cells with D2D}
\author{\IEEEauthorblockN{Arun Ramamurthy$^\dag$, Vanlin Sathya$^\dag$, Shrestha Ghosh$^*$, Antony Franklin$^\dag$ and Bheemarjuna Reddy Tamma$^\dag$}
\IEEEauthorblockA{$^*$Department of Computer Science and Technology,\\ Indian Institute of Engineering Science and Technology, Shibpur, India\\$^\dag$Department of Computer Science and Engineering,\\ Indian Institute of Technology Hyderabad, India}
{Email: [me11b005, cs11p1003]@iith.ac.in, ghosh\_shrestha@yahoo.co.in, [antony.franklin, tbr]@iith.ac.in.}
}

%\vspace{-0.5cm}
\maketitle
\begin{abstract}
The recent developments in full duplex (FD) communication promise doubling the capacity of cellular networks using self interference cancellation (SIC) techniques. FD small cells with device-to-device (D2D) communication links could achieve the expected capacity of the future cellular networks (5G). In this work, we consider joint scheduling and dynamic power algorithm (DPA) for a single cell FD small cell network with D2D links (D2DLs). We formulate the optimal user selection and power control as a non-linear programming (NLP) optimization problem  to get the optimal user scheduling and transmission power in a given TTI. Our numerical results show that using DPA gives better overall throughput performance than full power transmission algorithm (FPA). Also, simultaneous transmissions (combination of uplink (UL), downlink (DL), and D2D occur 80\% of the time thereby increasing the spectral efficiency and network capacity. %We also analyze results obtained for controlled D2D transmissions by varying the 
%allowable D2D. The average D2D throughput increases when more D2D transmission are allowed without altering the average throughputs in the DL and UL. 
\end{abstract}
\begin{IEEEkeywords}
5G, Small cell, FD, D2D communication, NLP.
\end{IEEEkeywords}
%\vspace{-0.2cm}
\section{Introduction}
The number of cellular user equipments (CUEs) demanding mobile data is increasing by leaps and bounds, be it for instant image uploading or live video streaming. Therefore, in order to address the bandwidth crunch, telecom operators have resorted to small cell deployments in urban areas. By employing intelligent traffic and radio resource management mechanisms, small cells like pico cells and femto cells can deliver seamless network connectivity together with high system capacity. Due to the availability of limited spectrum, reuse of spectrum is essential to increase the system capacity. Recently researchers have shown that full duplex (FD) small cells have huge potential to drastically improve the capacity by using the same radio resources for the uplink (UL) and the downlink (DL) communications simultaneously~\cite{goyal2014improving}. Another technique that can be used to improve the system capacity is by allowing the small cells to support device-to-device (D2D) communication between proximity devices~\cite{feng2013device}.

In FD communications~\cite{bharadia2013full}, a base station (BS) is capable of transmitting (in the DL to UEs) and receiving (in the UL from UEs) signals at the same time on the same channel. Such simultaneous transmissions and receptions lead to self-interference at the BSs. However, with the evolving self-interference cancellation (SIC)~\cite{goyal2014improving} techniques FD communications have challenged the legacy half duplex (HD) mechanism employed at the BSs. Theoretically the FD communication is capable of reducing the spectrum demand by half~\cite{nguyen2014spectral} as the BSs perform FD operation in each transmission time interval (TTI) in LTE/LTE-A. But FD operation cannot be executed in all TTIs \cite{sultan2014impact} due to prevailing interference conditions. FD operation is dependent on relative positions of the CUEs from the BS, channel quality, and SIC capability of the BS~\cite{goyal2014full}. The BS allows simultaneous transmissions only if the interference levels are within a threshold. 
Soon after the SIC limits became practical, small cells like femtos adopted FD communication. Thus the CUEs can now connect to FD femto BS (FD FBS).

Apart from FD communications, proximity user equipments (UEs) under a BS coverage can directly communicate with each other without routing their traffic through the BS using D2D communication. The two UEs which perform direct communication forming a D2D link (D2DL) effectively offload local traffic. This allows the network to serve more UEs as D2DLs using the same radio resources. Efficient D2D communication under Macro BS is a well studied problem. In ~\cite{bastug2014social}, D2D is applied in HD small cells to increase their capacity. 
\par D2D links in small cell networks serve as a viable solution for offloading traffic in LTE networks~\cite{ishii2013lte}. In~\cite{semiari2015context}, D2D radio interface exploits the proximity of the UEs and the small cell BS (SCBS) and the benefits of this offload solution is quantified by analysis and simulation. The D2D communication is between proximity UEs of users sharing a strong social relationship, so that, they can share the content of common interest. These common contents are cached in a serving UE and served to other UE's directly over the D2D link. However, the bandwidth is divided between the SCBS and the serving UE so that there is no mutual interference between them. 

%\begin{figure}[htb!]
%\begin{center}
%\includegraphics[width=6cm,height=5cm]{fulld.eps}
%\caption{An example of FD FBS with Cellular Users and one D2D Link.}
%\vspace{-0.3cm}
%\label{sysarch}
%\end{center}
%\end{figure}

%A small cell network which combines both D2D and FD techniques has a huge potential to improve the overall system performance. But, the interference in the system also increases as shown in Fig.~\ref{Architecture}. Hence, efficient power control and scheduling of users is essential for interference management. 
The main contributions of this paper are-
%\todo{Improve below two points this is very important}
\begin{itemize}
\item Analytical evaluation of the performance of FD small cell network with D2DLs and analysis of its relative performance with respect to HD small cell network with D2DLs. To the best of our knowledge, this is the first time both FD and D2D communications have been applied to small cells.
\item An dynamic power control algorithm (DPA) for joint user selection and power allocation which highlights the potential of FD small cell networks with D2DLs.
\item Analysis of system throughputs for different weightages of the D2DLs.
\end{itemize}

We provide a numerical analysis of the DPA to show the potential of FD small cell with D2DLs. In the best case, in a TTI, two legacy cellular communications (one UL and one DL) and a D2D communication will take place simultaneously by sharing the same radio resource. Numerical results show significant improvement in system throughput while using dynamic power control in DPA than using the full power algorithm (FPA) in a small cell FD network. 

The rest of paper is organized as follows. Section~\ref{SA} describes the assumptions in system model and utility function. Section~\ref{FPA} and Section~\ref{OPCA} describes the full power and dynamic power control algorithm, respectively. Section~\ref{exp} explains the experimental setup and numerical results with comparison of FPA and DPA, Effects of SIC values, Different weightage. Finally Section~\ref{con} concludes the work.

\section{System Model and Utility Funtion}\label{SA}
\begin{figure}[htb!] 
\begin{center} 
\includegraphics[width=8cm,height=5cm]{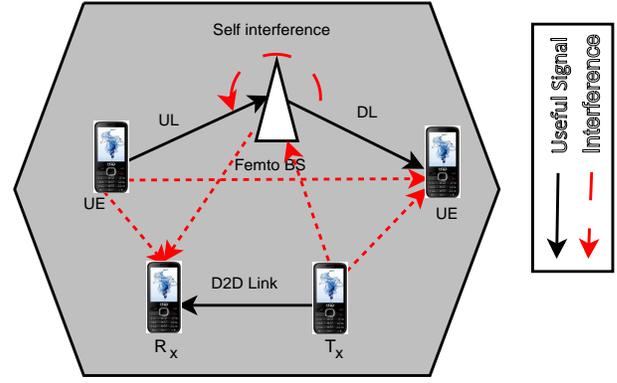}
%\vspace{0.3cm}
\caption{A FD FBS System with CUEs and D2DLs.}
%\vspace{-0.3cm}
\label{Architecture}
\end{center}
\end{figure}

We have modelled our system as shown in Fig.~\ref{Architecture} with a single femto BS (FBS) serving a set $\mathcal{C}$ of CUEs. CUEs have both UL and DL traffic demands. Further, FBS helps in establishing a set $\mathcal{L}$ D2DLs, (\emph{i.e.,} $2\times|\mathcal{L}|$ D2D UEs who always communicate over D2DLs). The FBS operates in hybrid mode (FD or HD) depending on the current network conditions. Theoretically the FBS can operate in FD mode in all TTIs thereby doubling the network capacity. But it is not practical due to the prevailing self-interference in the FD FBS and D2D interference. The UL CUE creates interference for the DL CUE and the receiver ($R_x$) of the D2DL. The $R_x$ of D2DL experiences interference from the FBS in the DL. The transmitter ($T_x$) in the D2DL creates interference for the FBS and the DL CUE. The CUEs are configured to operate only in the HD mode due to complexities involved in employing FD operations in the CUEs~\cite{goyal2014improving, goyal2014full}. We assume that each UE 
has full buffer traffic and require all available radio resources in all TTIs so that in a TTI at most one DL CUE, one UL CUE, and one D2DL may be scheduled. Although many D2DLs can be scheduled in a TTI, for simplicity, we considered only one D2DL to be active in a given TTI. Hence, in a TTI, the FBS may serve either \{two CUEs (one UL and one DL)\} or \{two CUEs (one UL and one DL) and a D2DL\} or \{one CUE (UL or DL)\} or \{one CUE (UL or DL) and a D2DL\}. But the model can be extended to allow multiple D2DLs per TTI which would further enhance the overall throughput. In our model we have assumed that a D2D UE is served in a single time slot, \emph{i.e.,}, a D2DL is formed between two proximity UEs if one of the UEs has the data which the other UE demands.
%\vspace{-0.4cm}
%\vspace{-0.2cm}

Since the FD mode is not active in all TTIs, the scheduler should opportunistically schedule CUEs in the FD mode whenever appropriate. Thus, our scheduler meets the typical criteria of most schedulers, \emph{i.e.,} maximizing the system throughput, maintaining a level of fairness among all the UEs. Our scheduler aims to maximize the system throughput by considering the logarithmic sum of the average data rate of all the UEs ($O^t$) at time $t$ as given by Eqn.~\eqref{SchObj}.
%~\cite{tse2005fundamentals}. 
%\vspace{-0.2cm}
\begin{equation} \label{SchObj}
%O^t = w_D \displaystyle\sum\limits_{i \in C} \log{\bar{R}_i^t} + w_U \displaystyle\sum\limits_{j \in C} \log{\bar{R}_j^t} + w_L \displaystyle\sum\limits_{k \in L} \log{\bar{R}_k^t}
O^t = \sum_{d\in\mathcal{C}} w_d\log{\bar{R}_d^t} + \sum_{u\in\mathcal{C}} w_u\log{\bar{R}_u^t} + \sum_{l\in\mathcal{L}} w_l\log{\bar{R}_l^t} 
\end{equation}
%\vspace{-0.2cm}
where $\bar{R}_{d}^t$, $\bar{R}_{u}^t$, and $\bar{R}_{l}^t$ are the average achieved data rates of DL CUE $d$, UL CUE $u$, and D2DL $l$ at a particular TTI $t$, respectively. The $w_d$, $w_u$, and $w_l$ are the throughput weightages of DL CUE $d$, UL CUE $u$, and D2DL $l$, respectively, which take values between 0 and 1. The average achieved data rate $\bar{R}_{x}^t$, at given TTI $t$, depends on the data rate of $x$ in the previous TTI (\emph{i.e.,} at time $t-1$) and the instantaneous data rate achieved at time $t$ is given by Eqn.~\eqref{AAR}.
%\vspace{-0.2cm}
\begin{equation}\label{AAR}
\bar{R}_{x}^t=\beta\bar{R}_{x}^{t-1}+\gamma\hat{R}_{x}^t
\end{equation}
where $x$ can be DL CUE, UL CUE, or D2DL. $\beta$ and $\gamma$ are weighing factors whose value lies between 0 and 1, and $\beta+\gamma = 1$. $\bar{R}_x^{t-1}$ is the average achieved data rate of CUE/D2DL $x$ at TTI $t-1$, and $\hat{R}_x^t$ is the instantaneous data rate of $x$ at $t$ which is calculated as shown in Eqn.~\eqref{IDR}.
\begin{equation}\label{IDR}
\hat{R}_x^t = B \times SE(SINR_x^t)
\end{equation}
where $B$ is the bandwidth assigned and $SE(SINR_x^t)$ is the spectral efficiency when the SINR of CUE/D2DL $x$ at TTI $t$ is $SINR_x^t$. SE can be measured using the CQI table in LTE/LTE-A~\cite{2013}. When a CUE/D2DL is not scheduled at $t$, the instantaneous throughput $\hat{R}_x$ is equal to zero. Since computing the throughput of the entire system in each TTI is cumbersome, we define a utility function derived from the system throughput. The value of the utility function depends on the UEs which are scheduled in a TTI. The FBS undertakes simultaneous transmissions if there is a UE combination whose utility value is higher than that of a single CUE. If DL CUE $i^*$, UL CUE $j^*$, and D2DL $k^*$ are scheduled in a TTI $t$, their utility is given in Eqn.~\eqref{UtilityValue}. The overall utility is nothing but the sum of downlink utility ($utility_d$), uplink utility ($utility_u$) and D2D utility ($utility_l$). Derivation of Eqn.~\eqref{UtilityValue} from Eqn.~\eqref{SchObj} is provided in the Appendix. 

\begin{scriptsize}
\begin{eqnarray} \label{UtilityValue} 
OverallUtility = \begin{cases}
			 w_{d^*}\big[\log{\big(\beta\bar{R}_{d^*}^{t-1}+\gamma\hat{R}_{d^*}^{t}\big)} - \log{\beta\bar{R}_{d^*}^{t-1}}\big] + \\
			 w_{u^*}\big[\log{\big(\beta\bar{R}_{u^*}^{t-1}+\gamma\hat{R}_{u^*}^{t}\big)} - \log{\beta\bar{R}_{u^*}^{t-1}}\big] + \\
			 w_{l^*}\big[\log{\big(\beta\bar{R}_{l^*}^{t-1}+\gamma\hat{R}_{l^*}^{t}\big)} - \log{\beta\bar{R}_{l^*}^{t-1}}\big]
			 \end{cases}
\end{eqnarray}
\end{scriptsize}

The estimation of the utility function is presented in Algorithm~\ref{UF_Alg}. It takes as input the selected combination of links ($d$, $u$, $l$), the transmit power of the corresponding source for each link ($p_{d}$, $p_{u}$, $p_{l}$). $G_{x \rightarrow y}$ is the channel gain between $x$ and $y$. The $l^{t}$ and $l^{r}$ are the transmitter and receiver of D2DL $l$. $N_d$, $N_u$, and $N_l$ are the DL, UL, and D2D noise, respectively. The algorithm returns the utility of the combination of UEs passed to it.
%\vspace{-0.2cm}
\begin{algorithm}[htb!] 
\caption{Utility Calculation} \label{UF_Alg}
%\textbf{Input:} $N_d$, $N_u$, $N_l$, $\beta$, $B$, $SIC$, $t$
%\hrule
\begin{algorithmic}[1]
\Function{UtilVal}{$\{d,p_d\},\{u,p_u\},\{l,p_l\}$}
\State $SINR_d^t \la \dfrac{G_{FBS \ra d}p_d}{N_d+G_{u \ra d}p_u+G_{l^t \ra d}p_l}$
\State $SINR_u^t \la \dfrac{G_{u \ra FBS}p_u}{N_u+G_{l^t \ra FBS}p_l+ \frac{p_d}{SIC}}$
\State $SINR_l^t \la \dfrac{G_{l^t \ra l^r}p_l}{N_l+G_{FBS \ra l^r}p_d+G_{u \ra l^r}p_u}$
\State $\hat{R}_{d}^t \la B \times SE(SINR_d^t)$
\State $\hat{R}_{u}^t \la B \times SE(SINR_u^t)$
\State $\hat{R}_{l}^t \la B \times SE(SINR_l^t)$
\State $Utility_d \la \log{\big(\beta\bar{R_d}^{t-1} + \gamma\hat{R}_{d}^t\big)} - \log{\beta\bar{R_d}^{t-1}}$
\State $Utility_u \la \log{\big(\beta\bar{R_u}^{t-1} + \gamma\hat{R}_{u}^t\big)} - \log{\beta\bar{R_u}^{t-1}}$
\State $Utility_l \la \log{\big(\beta\bar{R_l}^{t-1} + \gamma\hat{R}_{l}^t\big)} - \log{\beta\bar{R_l}^{t-1}}$
\State $OverallUtility \la Utility_d + Utility_u + Utiliy_l$
\State return $OverallUtility$ 
\EndFunction
\end{algorithmic} 
\end{algorithm}

In a TTI, the scheduler selects the combination of UEs (CUEs and/or D2D UEs) which has the highest utility. In order to compute the utility, the power values are determined as provided as an input to the utility function. Intuitively, we can state that the achieved system throughput is less when all transmitters transmit at their full (peak) power than when the power levels are controlled. Because, the level of interference at full power is higher than the case when power is controlled. We developed the DPA such that the algorithm performs joint user selection (including the D2DL) and power control. We then analyze the performance of DPA against FPA and HD mode to conclude how FD small cell network with D2DLs can increase network capacity.
%\vspace{-0.2cm}
\section{Full Power Algorithm (FPA)}\label{FPA} 
In FPA, all transmissions are carried out at the peak power level. It selects the combination of UEs, (\emph{i.e.,} DL, UL, and D2DL) whose utility is the maximum. Thus the objective of the scheduler (refer Eqn.~\eqref{FPA_equn}) is to find $d^*$, $u^*$, and $l^*$ which result in the maximum utility. 

%\vspace{-0.2cm}
\begin{scriptsize}
\begin{equation} \label{FPA_equn}
[d^*,u^*,l^*] = \argmax{\substack{i \in \mathcal{C} \cup \{0\};j \in \mathcal{C} \cup \{0\};\\k \in \mathcal{L} \cup \{0\};i \neq j}} \text{UtilVal}(\{i,P_{FBS}^{max}\},\{j,P_{CUE}^{max}\},\{k,P_{D2D}^{max}\})
\end{equation}
\end{scriptsize}
%\begin{equation}\label{eq6}
%\footnotesize
%(i^*, j^*, k^*)= \max_{_{\substack{0 \leq i \leq \lvert C \rvert;0 \leq j \leq \lvert C \rvert,\\ i \neq j; 0 \leq k \leq \lvert L \rvert}}}Utility(\{i, P_{FBS}^{max}\}, \{j, P_{U}^{max}\}, \{k, P_{L}^{max}\})
%\end{equation}
where, $P_{FBS}^{max}$, $P_{CUE}^{max}$, and $P_{D2D}^{max}$ are the maximum allowed transmission powers of the FBS, UL CUEs, and D2D transmitters, respectively. If only one CUE is selected (only UL or DL) then either $d^*$ or $u^*$ is zero. Similarly if no D2DL is selected, then $l^*=0$. $i\neq j$ ensures that the same UE cannot be selected for both UL and DL transmission in the same TTI.  The inefficiency of FPA lies in the fact that at full power transmission the interference level at FBS is high.

\section{Dynamic Power Control Algorithm}\label{OPCA}
In order to address the inefficiency of FPA, we propose Dynamic Power Control Algorithm (DPA) to be applied in every TTI. In this algorithm, UEs combination are chosen with maximum utility value. The scheduler in DPA determines the optimal transmit powers of a UE combination [$i$, $j$, $k$] by solving the following non-linear optimization model.
%\vspace{-0.2cm}
\begin{equation} \label{NLP_obj}
\underset{p_i,p_j,p_k}{maximize} \text{ UtilVal}(\{i,p_i\},\{j,p_j\},\{k,p_k\})
\end{equation}
%\vspace{-0.3cm}
subject to,
%\vspace{-0.5cm}
%\label{NLP_powconDL} \label{NLP_powconUL} \label{NLP_powconD2D} 
\begin{equation}\label{NLP_powconDL} 
p_i \leq P_{FBS}^{max};  
p_j \leq P_{CUE}^{max};  
p_k \leq P_{D2D}^{max}
\end{equation}
%\vspace{-0.3cm}
\begin{footnotesize}
\begin{equation}\label{NLP_sinrconDL}
SINR_i \geq SINR_i^{min};
SINR_j \geq SINR_j^{min};
SINR_k \geq SINR_k^{min}
\end{equation}
\end{footnotesize}
%\vspace{-0.5cm}
\begin{equation}
p_i, p_j, p_k \geq 0 \label{NLP_lb} 
\end{equation}
where, $SINR_i^{min},$ $SINR_j^{min},$ and $SINR_k^{min}$ is the minimum required SINR for DL CUE $i$, UL CUE $j$, and D2DL $k$. The NLP model maximizes the utility value (Eqn.~\eqref{NLP_obj}) for the UE combination $[i, j, k]$. Eqn.~\eqref{NLP_powconDL} ensures that the transmission power doesn't exceed the peak power. Eqn.~\eqref{NLP_sinrconDL} ensures that the SINR of the CUEs and D2DLs is above the minimum required SINR. Eqn.~\eqref{NLP_lb} specifies the lower bound.

After finding the optimal powers for each possible combination, it selects the combination which gives the highest utility as shown in Eqn.~\eqref{oPCA_equn}. 

%\vspace{-0.2cm}
\begin{scriptsize}
\begin{equation} \label{oPCA_equn}
[d^*,u^*,l^*] = \argmax{\substack{i \in \mathcal{C} \cup \{0\};j \in \mathcal{C} \cup \{0\};\\k \in \mathcal{L} \cup \{0\};i \neq j}} \text{UtilVal}(\{i,p_{ijk,i}^{*}\},\{j,p_{ijk,j}^{*}\},\{k,p_{ijk,k}^{*}\})
\end{equation}
\end{scriptsize}

where $p_{ijk,i}^{*}$, $p_{ijk,j}^{*}$, and $p_{ijk,k}^{*}$ denote the optimal transmission power of DL CUE $i$, UL CUE $j$, and D2DL $k$, respectively in the UE combination [$i,j,k$]. 

%\todo{Rewrite the below paragraph by saying "As the objective function is an NLP model is non-continuous ...hence we use pattern search algo...} 

\section{Experimental Setup and Numerical Results}\label{exp}
The optimization problem with the assumption of system model described in Section~\ref{SA} have been simulated using MATLAB~\cite{matlab}. As the objective function (Eqn.~\eqref{NLP_obj}) of the NLP model is non-continuous and non-differentiable, gradient based methods (steepest descent etc.) cannot be used to solve the model. Hence, we use \textit{Pattern Search Algorithm} in MATLAB to solve the NLP model. In the experimental setup, there are 10 CUEs who can transmit and receive data through FBS and 10 D2D UEs who can form 5 D2DLs. The CUEs and D2DLs are uniformly distributed in a rectangular region of dimensions $60m \times 50m$. The FBS is located in the center of the rectangle. The maximum length of a D2DL is set as $4m$. For a single small cell scenario, in our simulations, we put SINR requirement of 10.37 dB~\cite{goyal2014improving} (CQI Class 10) at the edge to calculate the required transmission power in both directions. Assuming the NLOS propagation (147.4+43.3 $log_{10}(R)$, distance (R) in km), 
the required FBS power ($P_{FBS}^{max}$) for 10.37 dB SINR is 1.78 dBm and the CUE ($P_{U}^{max}$) \& D2D ($P_{L}^{max}$) power is 0.78 dBm. The noise figures for DL, UL and D2D are 8, 9 and 8 dB respectively and the thermal noise density is -~174 dBm/Hz. We ran the simulation for 100 seconds with an average of 50 randomly generated scenarios and the system bandwidth is 10 MHz. The SIC values are varied from $65~dB$ to $105~dB$. We provide detailed analysis of the performance of DPA in a more challenging scenario created by lower SIC value (65 dB). We have kept the same throughput weightage for the CUEs and D2DLs, (\emph{i.e.}, $w_d$ = $w_u$ = $w_l$). We also obtained results for unequal weightage of UEs which are inline with the results presented here, and, are shown in the last section. 
%\vspace{-0.1cm}
%\begin{table}[htb!]\caption{Simulation parameters}
%\vspace{-0.2cm}
%\centering
%\begin{tabular}{p{4.8cm} p{3.2cm}}
%\hline
%\bfseries{Parameter}&\bfseries{Value}\\
%\hline
%%\hline
%Path Loss (dB) (distance (R) in km)& LOS: 89.5+16.9 $log_{10}(R)$\\
%	& NLOS: 147.4+43.3 $log_{10}(R)$\\
%%\hline
%Thermal Noise Density & -174 dBm/Hz\\
%%\hline
%DL, UL, and D2D noise figures  & 8, 9, and 8 dB\\
%%\hline
%Simulation time & 100 Seconds\\
%%\hline
%Bandwidth & 10 MHz\\
%\hline
%\end{tabular} 
%\label{tab1}
%\vspace{-0.5cm}
%\end{table}

\subsection{Comparison of FPA and DPA}
%\begin{figure}[htb!]
%\begin{center}
%\includegraphics[width=7cm]{ulTpExt.eps}
%\caption{Cumulative Distribution Function (CDF) of Average Throughput of UL users}
%\label{fig1}
%\end{center}
%\end{figure}

%Fig.~\ref{fig1} shows the CDF of UL, DL, and D2D throughput in FPA and DPA. DPA improves UL throughput by $27\%$ over FPA. In FPA, the self-interference is high as FBS transmits at full power. This high self-interference creates noise and therefore decreases the UL users throughput. However, as the DL user is chosen to be as far away as possible from the UL user, the DL users throughput in FPA is better than that of the UL users. The improvement in DL throughput (7\%) is not as high as that achieved for the UL throughput. This can be explained by the same reason that UL users are more affected by the transmitting power of the FBS than the DL users. Since in DPA the transmission power of the FBS is optimally chosen by the algorithm, there are instances when the FBS operates at a lower transmission power. The interference to the UL user though independent of the scheduled DL user (or the D2D user) is very much affected by the transmitting power of the FBS. As a result, the interference to the UL user (
%$P_{FBS}^{max}/SIC$ is greatly reduced in DPA when compared to FPA. The effect of reduced interference is reflected in high UL throughput improvements in DPA. Compared to FPA, the D2D throughput in DPA improves by $20\%$. 

\begin{figure*}[htb!]
\minipage{0.32\textwidth}
% % \epsfig{width=9cm,figure=table.eps}
\includegraphics[width=6.3cm,height=4.5cm]{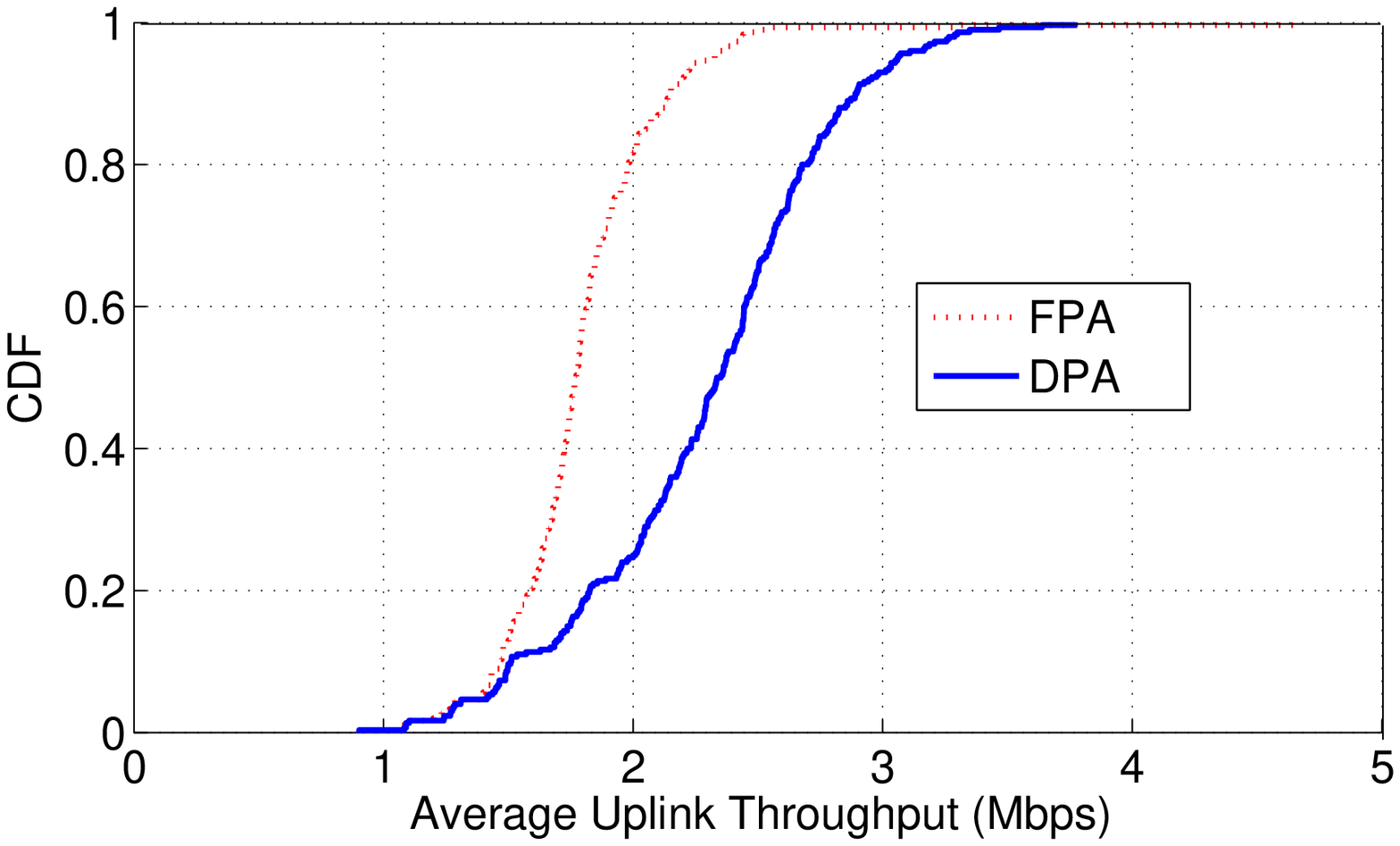}
\caption{Cumulative Distribution Function (CDF) of Average Throughput of UL CUEs}
\label{fig1}
\endminipage\hfill
~
\minipage{0.32\textwidth}
% % \epsfig{width=9cm,figure=table.eps}
\includegraphics[width=6.3cm,height=4.5cm]{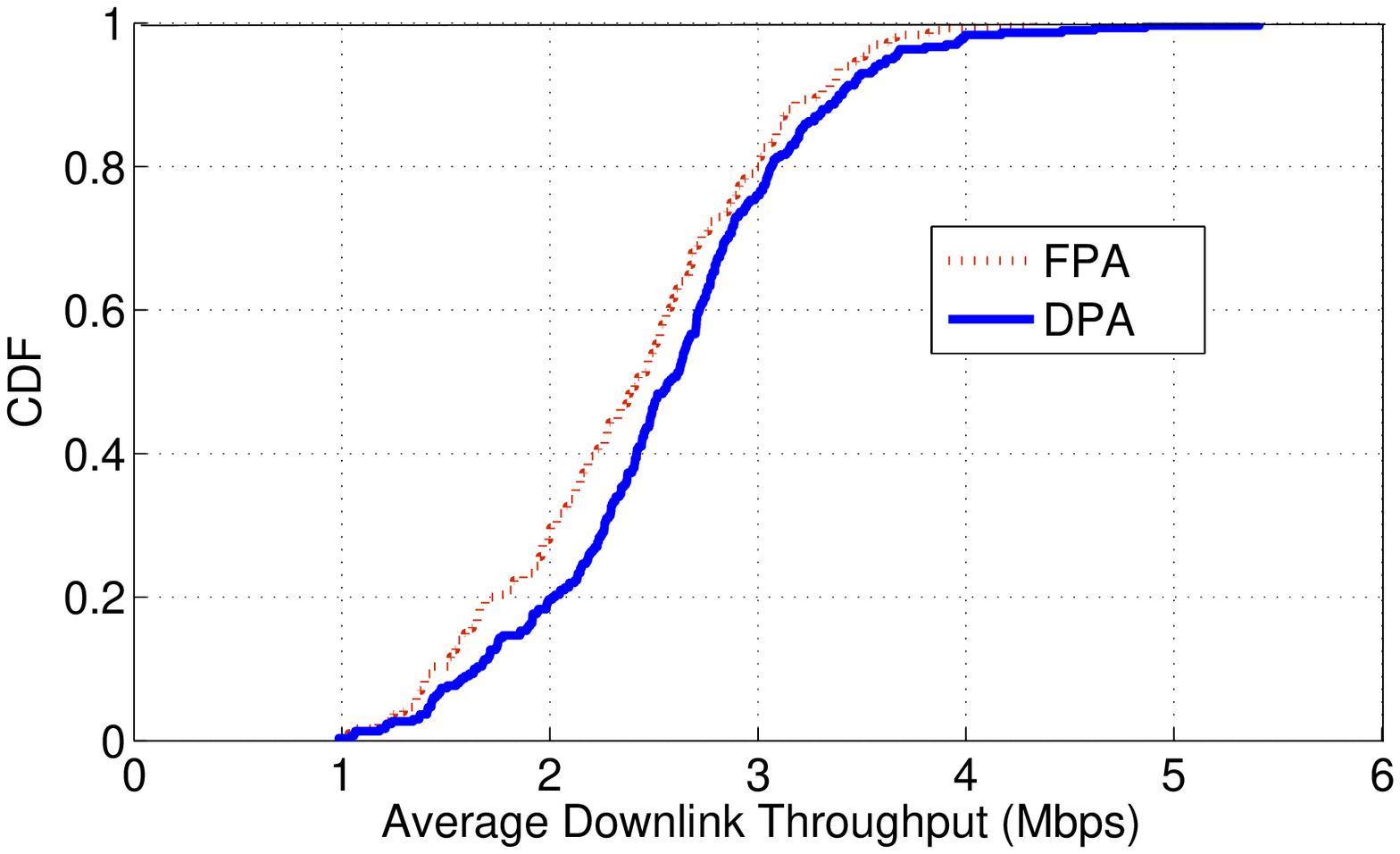}
\caption{Cumulative Distribution Function (CDF) of Average Throughput of DL CUEs}
\label{fig2}
%\vspace{-0.5cm}
\endminipage\hfill
~
\minipage{0.32\textwidth}
\includegraphics[width=6.3cm,height=4.5cm]{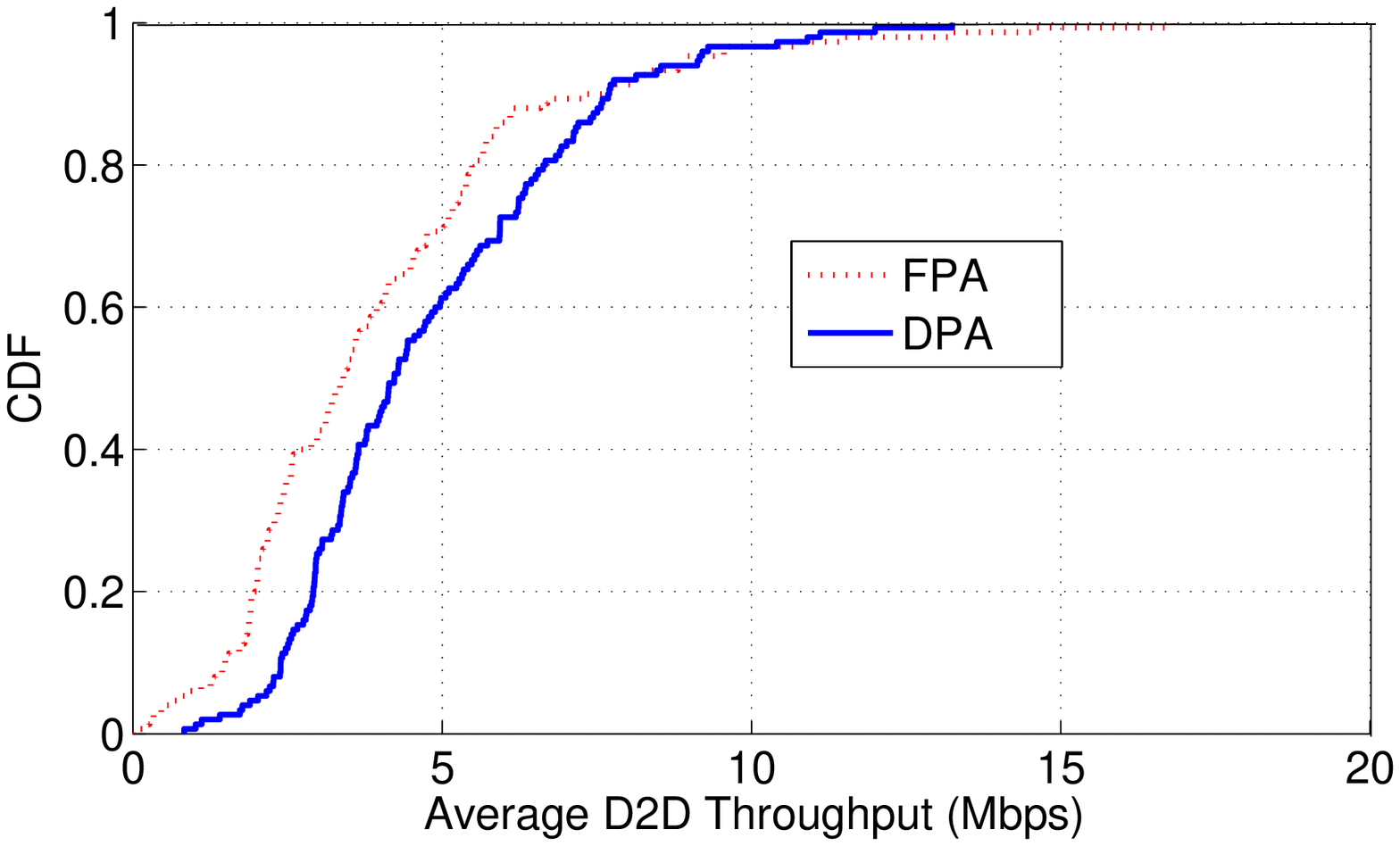}
\caption{Cumulative Distribution Function (CDF) of Average Throughput of D2D UEs}
\label{fig3}
\endminipage\hfill
\end{figure*}

\begin{figure}[htb!]
\begin{center}
%\vspace{-0.5cm}
\includegraphics[width=9cm,height=3cm]{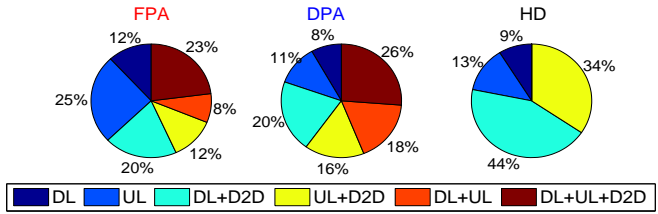}
%\vspace{-0.2cm}
\caption{Percentage Distribution of Different UE Combination.}
%\vspace{-0.5cm}
\label{fig4}
\end{center}
\end{figure}

%\todo{avoid shrinking vspace}
We compare CDF of average throughput of the cell obtained for the CUEs and D2DLs in DPA with respect to FPA and HD mode as shown in Fig.~\ref{fig1} with SIC value set to 65 dB. 
%\vspace{-0.1cm}
%\begin{figure}[htb!]
%\begin{center}
%%\vspace{-0.4cm}
%\includegraphics[width=8.3cm,height=4.2cm]{Results/TpExt12.eps}
%%\vspace{-0.2cm}
%\caption{CDF of Average Achieved Throughput in DL, UL, and D2D.}
%	%\vspace{-0.2cm}
%\label{fig1}
%\end{center}
%\end{figure}
%\vspace{-0.1cm}
In FPA, the DL CUE is chosen as far away as possible from any UL CUE or D2DL. However, the FBS transmits at the peak power independent of the DL UEs scheduled. This leads to high self-interference which results in lower UL throughput in FPA than the DL throughput. In DPA the transmission power of the FBS is optimally chosen, so, there are instances where the FBS operates at a lower transmit power (\emph{e.g.}, when a DL CUE is closer to the FBS). $P_{FBS}^{max}/SIC$ is greatly reduced in DPA rendering a 27\% improvement in the UL throughput and 7\% in DL throughput over that obtained in FPA. Since the UL CUEs are more affected than DL CUEs/D2D UEs in FPA, their improvement is higher. D2D throughput is higher than the UL and DL throughputs due to the shorter distance between the D2D UEs in a D2DL. Compared to FPA, the D2D throughput in DPA increases by 20\% because of the increased D2D opportunities in DPA than FPA (explained through Fig.~\ref{fig4}). The D2D throughput in HD mode is higher than FBS 
because of the absence of simultaneous UL and DL transmissions (refer Fig.~\ref{fig4}) and controlled transmission power of the FBS. The UL and DL throughputs can be improved by setting a higher weightage ($w_D$ and $w_U$) in Eqn.~\eqref{SchObj}. From Fig.~\ref{fig1}, we can also conclude that the DPA improves over HD mode in terms of UL and DL throughputs.
%The running time of DPA is of the order of few seconds whereas the scheduling decision needs to be taken in the order of few milliseconds. So, DPA is not practical. 
% Also, the running time of H-PCA is of the order of few milliseconds, which makes it practical. 
%\begin{figure}[htb!]
%\begin{center}
%\includegraphics[width=7cm]{dlTpExt.eps}
%\caption{Cumulative Distribution Function (CDF) of Average Throughput of DL users}
%\label{fig2}
%\end{center}
%\end{figure}

%\begin{figure}[htb!]
%\begin{center}
%\includegraphics[width=7cm]{d2dTpExt.eps}
%\caption{Cumulative Distribution Function (CDF) of Average Throughput of D2D users}
%\label{fig3}
%\end{center}
%\end{figure}

%\begin{table*}[htb!]
%\caption{simulation parameters}
%\centering
%\begin{tabular}{|p{2cm}| p{3.5cm}| p{3.5cm}| p{3.5cm}| p{3.5cm}|}
%\hline
%\bfseries{Algorithm} & \bfseries{Data/Joule(Gb/J)} & \bfseries{Increase in DL Throughput when compared to FPA} & \bfseries{Increase in UL Throughput when compared to FPA} & \bfseries{Increase in D2D Throughput when compared to FPA}\\
%\hline\hline
%FPA &\hspace{1cm}0.174 & \hspace{1cm}- & \hspace{1cm}- & \hspace{1cm}-\\
%\hline
%DPA &\hspace{1cm}0.313 & \hspace{1cm}$7.28\%$ & \hspace{1cm}$27.63\%$ & \hspace{1cm}$18.40\%$\\
%\hline
%SDPA &\hspace{1cm}0.232 & \hspace{1cm}$1.74\%$ & \hspace{1cm}$24.92\%$ & \hspace{1cm}$9.55\%$\\
%\hline  
%\end{tabular} 
%\label{tab2}
%\end{table*}

%\vspace{-0.1cm}
%Fig.~\ref{fig1} shows the achieved D2D throughput in the two algorithms.  As the power is moderated, more D2D links are eligible to be scheduled in a given TTI. 

%In a particular TTI, the different possible user combinations are a DL, an UL, DL and D2D, UL and D2D, DL and UL, and DL, UL and D2D. 
In Fig.~\ref{fig4}, the pie charts give the percentage distribution of different UE combinations for $100$ seconds of simulation time. The percentage of simultaneous transmissions in DPA ($80\%$) is higher than that of FPA ($62\%$) which implies effective reuse of the spectrum for optimally controlled transmit power levels of the nodes (FBS, CUEs, and D2DL). The D2D transmissions, which occur for a total of 80\% of a TTI in DPA, is comparable to $77\%$ in HD mode, both being greater than that in FPA ($55\%$). While the D2DL experience interference from atmost two CUEs in a TTI in DPA, there is only one interfering CUE in HD which explains the better D2D throughput in HD mode. We also found that Data/Joule (measured in $Gb/J$) in DPA ($0.319Gb/J$) is 86.5\% higher than FPA ($0.171 Gb/J$) and 6.3\% higher than HD mode ($0.300 Gb/J$).

\begin{figure}[htb!]
\begin{center}
%\vspace{-0.3cm}
\includegraphics[width=9cm,height=5cm]{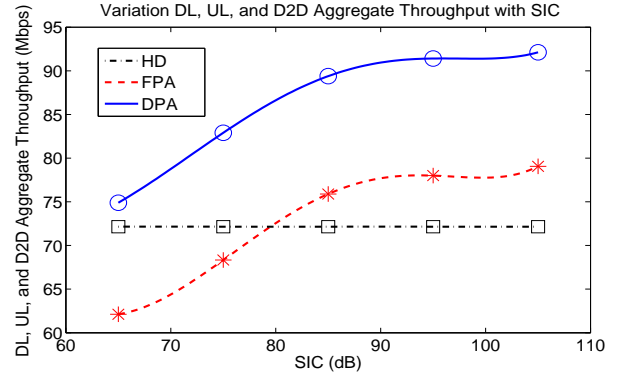}
%\vspace{-0.2cm}
\caption{Aggregate System Throughput vs SIC.}
%\vspace{-0.3cm}
\label{fig5}
\end{center}
\end{figure}
%\vspace{-0.3cm}
\subsection{Effect of SIC Values}
At low SIC values, say 65 dB, the hardware costs for implementing SIC mechanism is less. However, performing power control at lower SIC values becomes challenging. In Fig.~\ref{fig5} shows the variation of the aggregate system throughput for a range of SIC values (65 dB to 105 dB). The FD mode has an upper hand over the HD mode, because, increasing SIC values improves the system throughput in the FD mode. The changing SIC values does not affect the curve for HD mode because there are no simultaneous transmissions.

At higher SIC values (above $95 dB$), the self-interference becomes negligible and the improvement in throughput becomes less prominent. Therefore, at SIC values greater than 90 dB, the aggregate throughput curves becomes flatter. DPA gives the aggregate throughput higher than both the HD mode and FPA. With controlled transmission power, instances of simultaneous transmissions (a single or two CUEs with a D2DL transmission) increase hence, increasing the overall system performance.

The difference between the aggregate throughputs obtained from Fig.~\ref{fig5} at 105 dB and that at 65 dB is 17 Mbps in DPA. Thus, we can say that low SIC values does not greatly affect the performance of DPA.
\begin{figure*}[htb!]
\minipage{0.32\textwidth}
% % \epsfig{width=9cm,figure=table.eps}
\includegraphics[width=6.3cm,height=5cm]{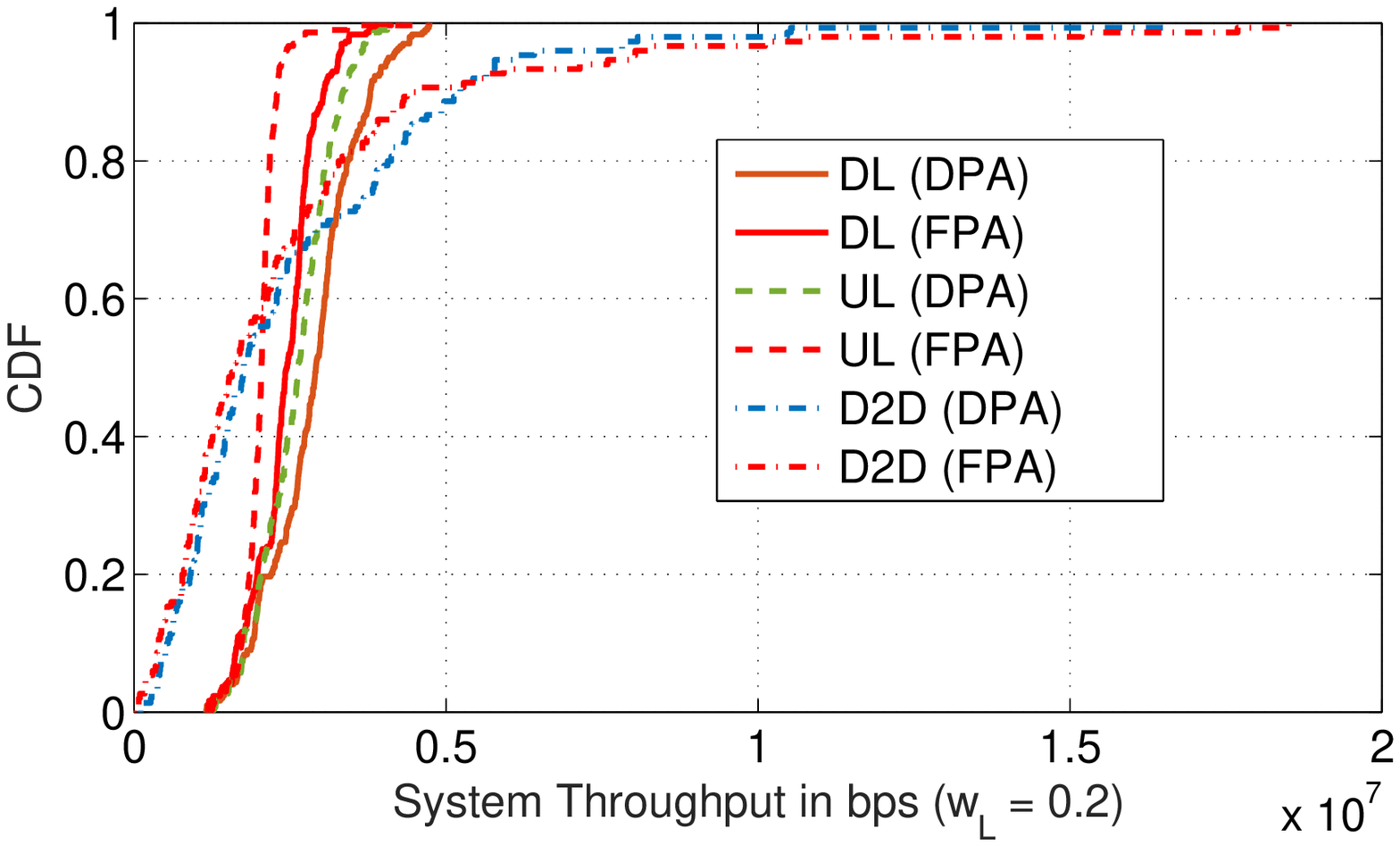}
\caption{Throughput with Weightage = 0.2 for 65 dB}
\label{W2}
\endminipage\hfill
~
\minipage{0.32\textwidth}
% % \epsfig{width=9cm,figure=table.eps}
\includegraphics[width=6.3cm,height=5cm]{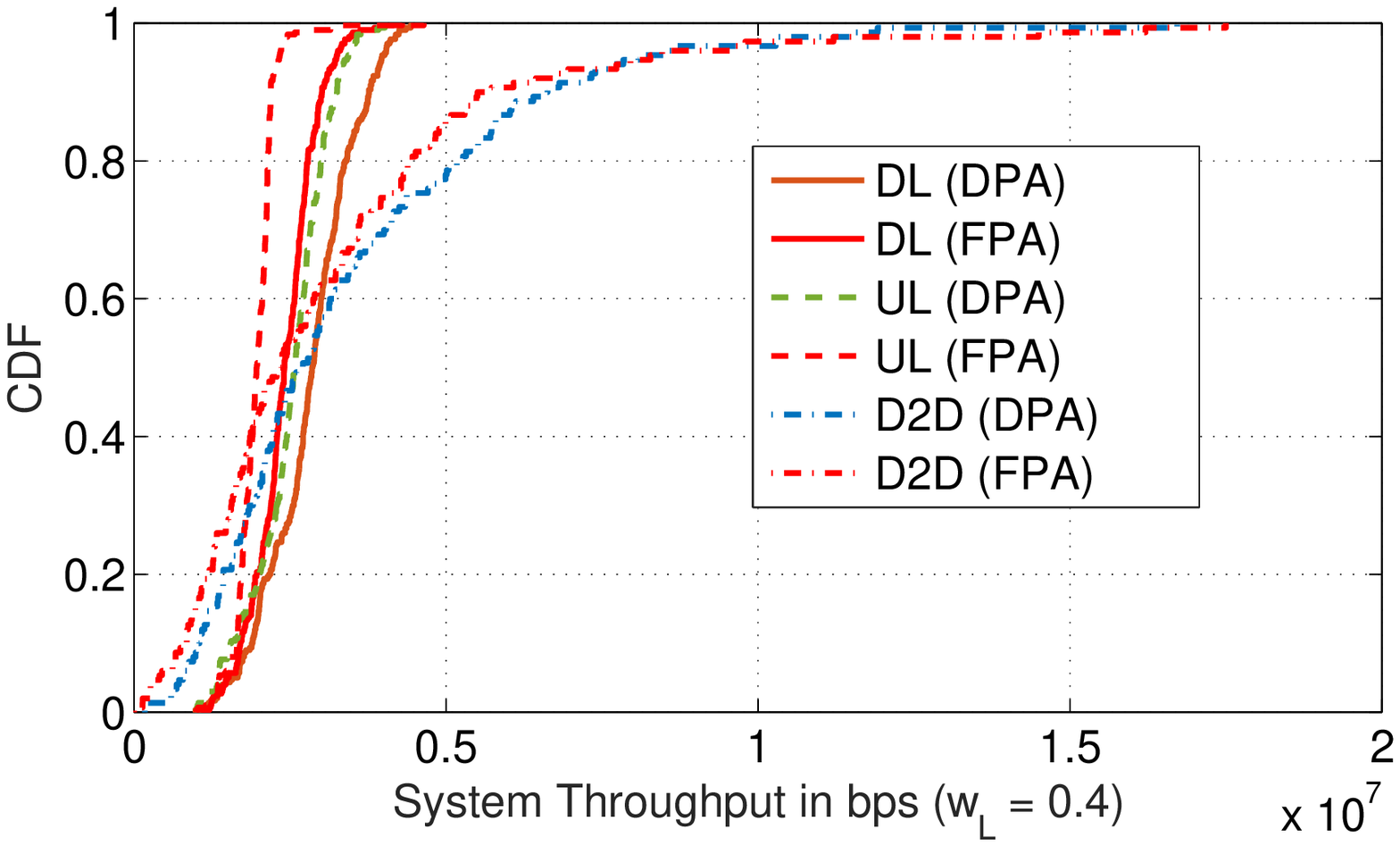}
\caption{Throughput with Weightage = 0.4 for 65 dB}
\label{W4}
%\vspace{-0.5cm}
\endminipage\hfill
~
\minipage{0.32\textwidth}
\includegraphics[width=6.3cm,height=5cm]{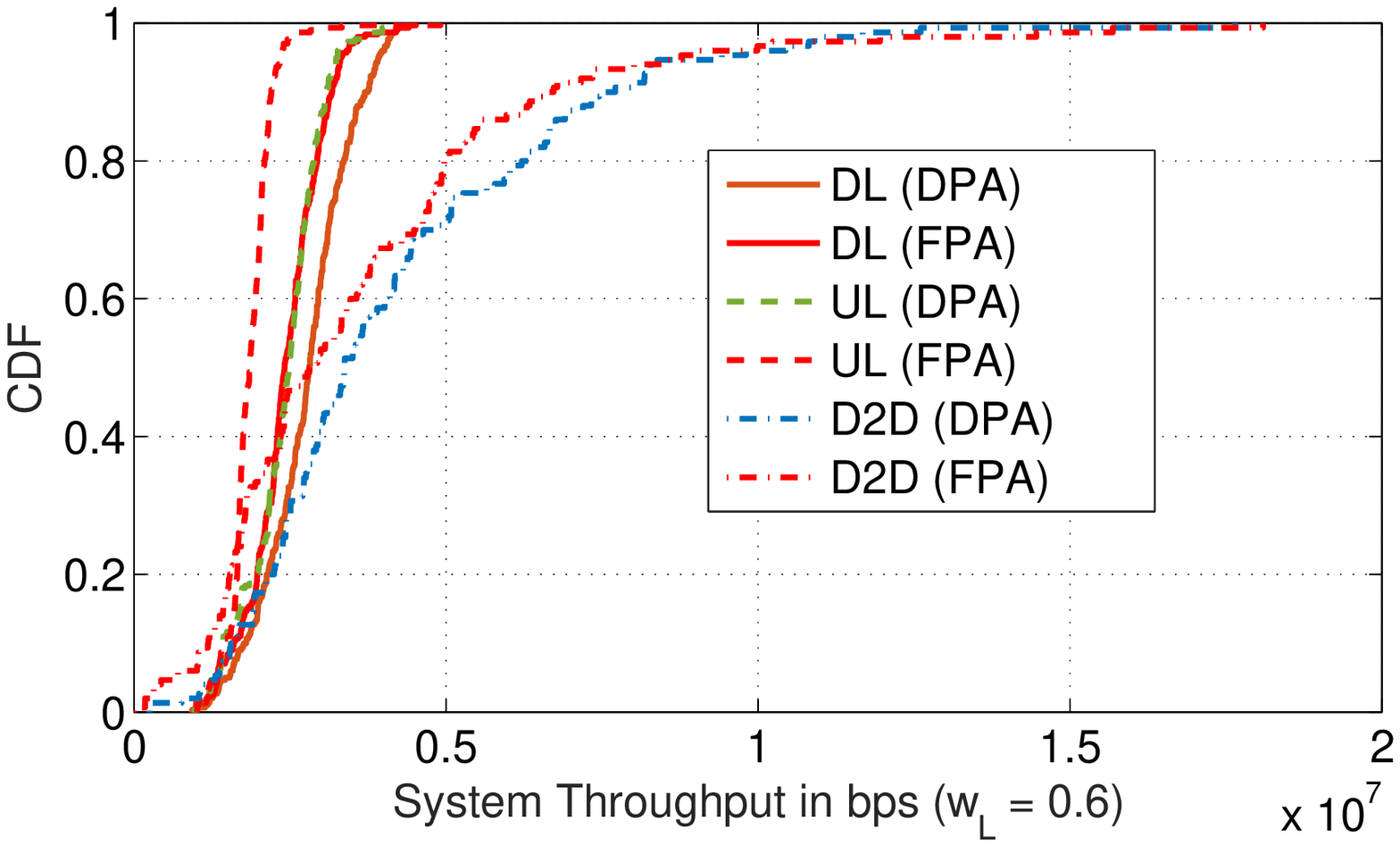}
\caption{Throughput with Weightage = 0.6 for 65 dB}
\label{W6}
\endminipage\hfill
\end{figure*}

%\begin{figure}[htb!]
%\begin{center}
%\vspace{-0.2cm}
%\includegraphics[width=8cm]{dit.eps}
%\vspace{-0.3cm}
%\caption{Decrease in Aggregate Throughput}
%\vspace{-0.3cm}
%\label{fig6}
%\end{center}
%\end{figure}
\begin{figure*}[htb!]
\minipage{0.5\textwidth}
% % \epsfig{width=9cm,figure=table.eps}
\includegraphics[width=8cm,height=5cm]{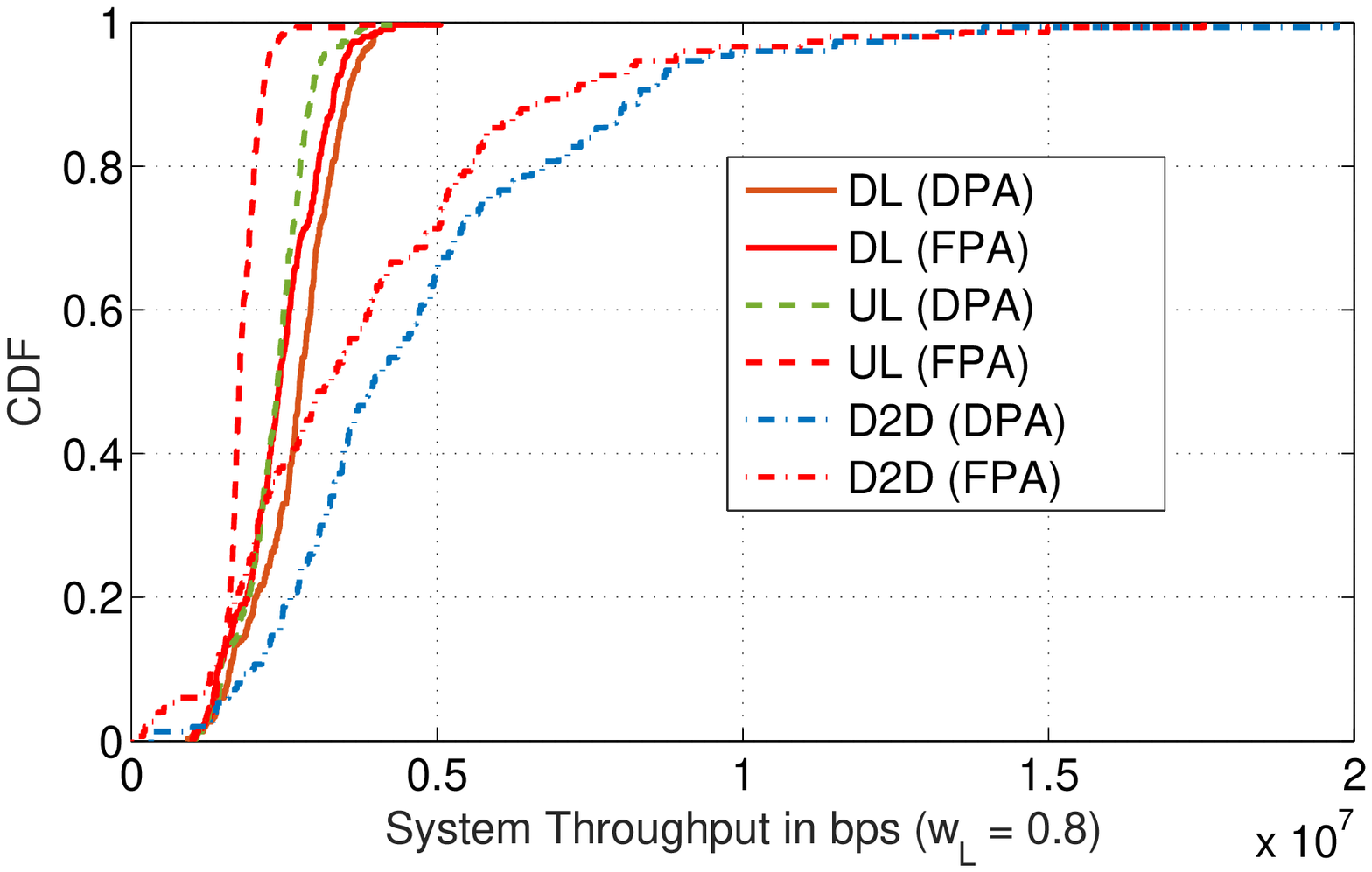}
\caption{Throughput with Weightage = 0.8 for 65 dB}
\label{W8}
\endminipage\hfill
~
\minipage{0.5\textwidth}
\includegraphics[width=8cm,height=5cm]{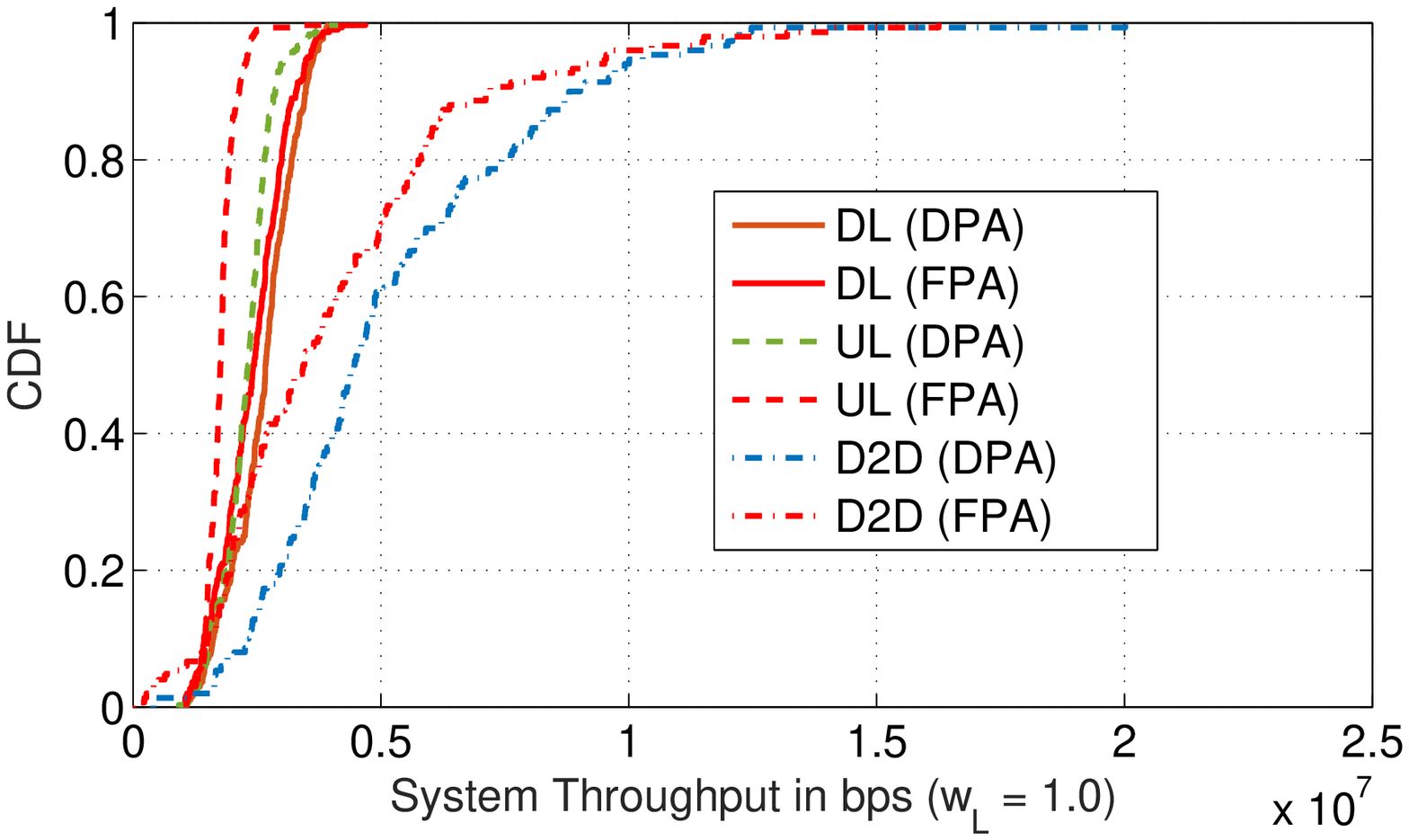}
\caption{Throughput with Weightage = 1 for 65 dB}
\label{W10}
\endminipage\hfill
%\vspace{-0.5cm}
\end{figure*}

\subsection{Different Weightage}
%\todo{write more elaborately and also say why we analyzed different weightage}
We derived the overall utility used by the scheduler to schedule transmissions in the UL, DL and D2DL (refer Eqn.~\eqref{UtilityValue}). The weightages $w_d$, $w_u$ and $w_l$ determines the importance of DL CUEs, UL CUEs and D2D UEs, respectively. Previously we considered all the UEs to be equally important ($w_d=w_u=w_l=1$). However, in real scenarios, the number of CUEs is much more that the number of D2DLs. Therefore, the weightage of D2D UEs may be kept at a lower value. By analyzing traffic conditions and user demand, a network operator may reduce $w_l$ when the D2D demand is low or vice versa.
%\begin{figure}
%\begin{center}
%%\minipage{0.5\textwidth}
%% % \epsfig{width=9cm,figure=table.eps}
%\includegraphics[width=8cm,height=6cm]{W10.eps}
%\caption{Throughput with Weightage = 1 for 65 dB}
%\label{W10}
%%\vspace{-0.5cm}
%\end{center}
%\end{figure}
In this section, we analyze the performance of DPA for different values of $w_l$. The number of D2D transmissions in the system is controlled by changing the value of $w_l$. Lower the value of $w_l$ lesser are the D2D transmissions that are carried out in a TTI. Figs.~\ref{W2}, \ref{W4}, \ref{W6}, \ref{W8}, \ref{W10} show the CDF plots of the average throughput in the DL, UL and D2D for $w_l \in \{0.2,0.4,0.6,0.8,1\}$ at 65 dB SIC, respectively. From the CDF plots we can observe that changing $w_l$ does not greatly affect the average throughput in the UL and DL. Here, increasing the number of D2D transmission does not reduce the UL \& DL throughput as the scheduling algorithm selects the UL \& DL such a way that the D2D will not interfere with them. As $w_l$ increases, the CDF curve of the average D2D throughput shifts to the right, \emph{i.e.,} the average D2D throughput for 60\% of the D2DLs increases from about 2Mbps for $w_l = 0.2$ to 5Mbps for $w_l = 1$.

%\todo{remove citation to MATLAB from References}
%\vspace{-0.1cm}
\section{Conclusions}\label{con}
In this paper, we showed the potential of FD small cell networks with opportunistic D2D communication by formulating the user selection and power control in a single cell as an optimization problem. Our numerical results showed that by using DPA, DL, UL, and D2DL throughputs can be improved by 27\%, 7\%, and 20\%, respectively compared to the FPA. We also analyse results obtained for different values of $w_l$. We would like to extend the proposed model in this paper to a heterogeneous network with multiple small cells inside a Macro cell in future. Also we would develop a heuristic algorithm for user selection and dynamic power control to be used in a practical network scenarios.\\
\noindent{\hspace{3cm}\textbf{Appendix}}\\ \\
\label{APDX} 
Let the DL CUE, UL CUE, and D2DL scheduled in TTI $t$ be $d^*$, $u^*$, and $l^*$, respectively. The utility value (Eqn.~\eqref{UtilityValue}) for the chosen combination can be derived from Eqn.~\eqref{SchObj} as follows,
%\begin{footnotesize}
\begin{eqnarray*}
O^t &=&  \sum_{d\in\mathcal{C}\bs d^*}w_d\log{\bar{R}_d^t} + w_{d^*}\log{\bar{R}_{d^*}^t} + \\
	&& \sum_{u\in\mathcal{C}\bs u^*}w_u\log{\bar{R}_u^t} + w_{u^*}\log{\bar{R}_{u^*}^t} + \\ 
	&& \sum_{l\in\mathcal{L}\bs l^*}w_l\log{\bar{R}_l^t} + w_{l^*}\log{\bar{R}_{l^*}^t} \\
	&=&  \sum_{d\in\mathcal{C}\bs d^*}w_d\log{\beta\bar{R}_d^{t-1}} + w_{d^*}\log{\big(\beta\bar{R}_{d^*}^{t-1}+\gamma\hat{R}_{d^*}^{t}\big)} + \\
	&& \sum_{u\in\mathcal{C}\bs u^*}w_u\log{\beta\bar{R}_u^{t-1}} + w_{u^*}\log{\big(\beta\bar{R}_{u^*}^{t-1}+\gamma\hat{R}_{u^*}^{t}\big)} + \\ 
	&& \sum_{l\in\mathcal{L}\bs l^*}w_l\log{\beta\bar{R}_l^{t-1}} + w_{l^*}\log{\big(\beta\bar{R}_{l^*}^{t-1}+\gamma\hat{R}_{l^*}^{t}\big)} \\
	&=& \sum_{d\in\mathcal{C}}w_d\log{\beta\bar{R}_d^{t-1}} + w_{d^*}\big[\log{\big(\beta\bar{R}_{d^*}^{t-1}+\gamma\hat{R}_{d^*}^{t}\big)} - \log{\beta\bar{R}_{d^*}^{t-1}}\big]  + \\
	&& \sum_{u\in\mathcal{C}}w_u\log{\beta\bar{R}_u^{t-1}} + w_{u^*}\big[\log{\big(\beta\bar{R}_{u^*}^{t-1}+\gamma\hat{R}_{u^*}^{t}\big)} - \log{\beta\bar{R}_{u^*}^{t-1}}\big]  + \\ 
	&& \sum_{l\in\mathcal{L}}w_l\log{\beta\bar{R}_l^{t-1}} + w_{l^*}\big[\log{\big(\beta\bar{R}_{l^*}^{t-1}+\gamma\hat{R}_{l^*}^{t}\big)} - \log{\beta\bar{R}_{l^*}^{t-1}}\big] \\
	&=& \sum_{d\in\mathcal{C}}w_d\log{\bar{R}_d^{t-1}}+\sum_{u\in\mathcal{C}}w_u\log{\bar{R}_u^{t-1}}+\sum_{l\in\mathcal{L}}w_l\log{\bar{R}_l^{t-1}}+\\
	&& \Big[\sum_{d\in\mathcal{C}}w_d+\sum_{u\in\mathcal{C}}w_u+\sum_{l\in\mathcal{L}}w_l\Big]\log{\beta}+\\
	&& w_{d^*}\big[\log{\big(\beta\bar{R}_{d^*}^{t-1}+\gamma\hat{R}_{d^*}^{t}\big)} - \log{\beta\bar{R}_{d^*}^{t-1}}\big]+\\
	&& w_{u^*}\big[\log{\big(\beta\bar{R}_{u^*}^{t-1}+\gamma\hat{R}_{u^*}^{t}\big)} - \log{\beta\bar{R}_{u^*}^{t-1}}\big]+ \\
	&& w_{l^*}\big[\log{\big(\beta\bar{R}_{l^*}^{t-1}+\gamma\hat{R}_{l^*}^{t}\big)} - \log{\beta\bar{R}_{l^*}^{t-1}}\big] \\
	&=& O^{t-1} + \Big[\sum_{d\in\mathcal{C}}w_d+\sum_{u\in\mathcal{C}}w_u+\sum_{l\in\mathcal{L}}w_l\Big]\log{\beta} + \\
	&& w_{d^*}\big[\log{\big(\beta\bar{R}_{d^*}^{t-1}+\gamma\hat{R}_{d^*}^{t}\big)} - \log{\beta\bar{R}_{d^*}^{t-1}}\big]+\\
	&& w_{u^*}\big[\log{\big(\beta\bar{R}_{u^*}^{t-1}+\gamma\hat{R}_{u^*}^{t}\big)} - \log{\beta\bar{R}_{u^*}^{t-1}}\big]+\\
	 && w_{l^*}\big[\log{\big(\beta\bar{R}_{l^*}^{t-1}+\gamma\hat{R}_{l^*}^{t}\big)} - \log{\beta\bar{R}_{l^*}^{t-1}}\big]
\end{eqnarray*}
%\end{footnotesize}
Since $O^{t-1},$ $\beta,$ $w_d,$ $w_u,$ and $w_l$ are constant at TTI $t$, $O^t$ depends only on the remaining terms in the above equation. Therefore, for a given UE combination $[d^*, u^*, l^*]$ utility is calculated as shown in Eqn.~\eqref{UtilityValue}.

\nocite{*}
%\vspace{-0.4cm}
\bibliographystyle{ieeetr}
\bibliography{IEEEabrv,references}
\end{document}